\documentclass[10]{article}

\usepackage{cite,amsmath,amssymb,graphicx,psfrag,subfigure}

\begin{document}

\title{Mean Field Models of Message Throughput in Dynamic Peer-to-Peer Systems\footnote{This work was in part funded by the Australian Research Council, ARC
Discovery Project, DP0451936.}}


\author{Aaron~Harwood and~Olga~Ohrimenko\\
Department of Computer Science and Software Engineering\\
The University of Melbourne, AUSTRALIA\\
\tt{\{aharwood,olgao\}@csse.unimelb.edu.au}}

\maketitle

\begin{abstract}
The churn rate of a peer-to-peer system places direct limitations on the rate
at which messages can be effectively communicated to a group of peers. These
limitations are independent of the topology and message transmission latency.
In this paper we consider a peer-to-peer network, based on the Engset model,
where peers arrive and depart independently at random. We show how the arrival
and departure rates directly limit the capacity for message streams to be
broadcast to all other peers, by deriving mean field models that accurately
describe the system behavior. Our models cover the unit and more general $k$
buffer cases, i.e. where a peer can buffer at most $k$ messages at any one
time, and we give results for both single and multi-source message streams.  We
define coverage rate as peer-messages per unit time, i.e. the rate at which a
number of peers receive messages, and show that the coverage rate is limited by
the churn rate and buffer size. Our theory introduces an Instantaneous Message
Exchange (IME) model and provides a template for further analysis of more
complicated systems. Using the IME model, and assuming random processes, we
have obtained very accurate equations of the system dynamics in a variety of
interesting cases, that allow us to tune a peer-to-peer system. It remains to
be seen if we can maintain this accuracy for general processes and when
applying a non-instantaneous model. 
\end{abstract}

\section{Introduction}

Fundamentally, in a peer-to-peer (P2P) network, messages can only be exchanged
between peers that are \emph{online} state. If a peer is not online then
it is \emph{offline} and no messages are exchanged with a peer
while it is offline.  When a peer is online then it is available to send and
receive messages from other peers. By considering each peer to be in one of
these two
states and by examining the frequency of state changes over all peers, we
describe the \emph{churn} of the P2P network. A high churn means a high
frequency of state changes and vice versa. We can bound the total number of
peers, in which case a given peer will continue to alternate between states
over time, or we can allow the number of peers to be infinite and consider a
finite number of online peers, in which case a given peer may never return to
the online state after becoming offline.  We call these cases the \emph{finite}
and \emph{infinite} population models respectively. In both cases there is an
expected number of peers that are online, at any given time, when the P2P
network is in equilibrium. In this work we focus on P2P networks with a bounded
number of peers because we are interested in how the churn affects the message
dissemination capacity of the network; though we make some remarks about the
infinite population model for interest.
 
Messages may be generated by users, application processes, data sources or as
the result of control traffic, e.g. stabilization or response to changing
network conditions, between peers. Using the finite population model we can
assume in this work that messages are to be disseminated to all other peers,
including to those that may happen to be offline at the time the message was
generated. In these circumstances we naturally ask for the time taken for a
message to be received by all (or a fraction) of the other peers with the
understanding that, as peers change between the offline and online states, all
peers may or may not eventually receive the message in question.

As one example, consider a query message that originates at a peer in an
unstructured P2P network; where each peer in the network may have some data
that is relevant to the query. In a basic network the query message is flooded
by each peer evaluating the query, forwarding the message to other peers and
possibly responding to the query. Typically each peer that receives the query
will delete it after consideration. In the finite population model we ask for
the fraction of total peers that received the query message, what we call the
query message \emph{coverage}. In the infinite population model we would ask
for the absolute number of peers that received the message. Intuitively, the
peers that receive the query message are those peers that were online while the
query message was being flooded. We can refine our intuition by considering
those peers that changed state, either from offline to online or vice versa,
during the time that the query message was being flooded. 

In general, messages may be buffered rather than being deleted immediately
after processing. If a message is buffered indefinitely by at least one peer
then (for a bounded number of peers) the message will eventually be received by
all peers, i.e. the coverage will be 100\%; this is the case for infinite sized
buffers. For a large volume of messages, e.g. consider when all peers in the
network are frequently generating queries, and there are limited resources
available at the peers, e.g. mobile devices versus desktops, it becomes more
practical to place a limit on the buffer size. In this case a message is
buffered only for some time and some peers may not receive the message at all.
In this work we examine how the size of the buffer relates to the coverage. To
do so we also consider the message rate.

\subsection{Instantaneous message exchange}

To clearly examine the affect from churn and finite buffer size we first
eliminate any affects from the sub-communication system (i.e. the Internet in
most cases) by allowing any number of message transmissions between online
peers to take place instantly; we refer to this as the \emph{instantaneous message
exchange} (IME) model. In this case, e.g., a message that is generated
at a peer that is online is instantly communicated to all other peers that are
also online at that time of message generation. While this generally is an
unrealistic allowance, it allows us to model how churn and finite buffer
size places limits on message coverage \emph{regardless} of the sub-communication
system's ideal performance. 

From another perspective The IME model is applicable when churn is sufficiently
low relative to the message propagation time through the network; in the
sense that the network appears to be static from the perspective of a single
message propagation. 

Aspects such as bandwidth and latency in the sub-communication system lead only
to further limitations; i.e. in this work we are concerned with how churn
places a limit on the message coverage and we make ideal or best case
assumptions about other aspects, which can otherwise only lead to the message
coverage being further limited.

%

%

\subsection{Related work}

To the best of our knowledge there are currently no results that have proposed
the IME model as a starting point. The dynamics of the IME model can be
analyzed by taking an epidemic information dissemination
approach~\cite{citeulike:292941}, sometimes called randomized rumor
spreading~\cite{karp}.  Based on analysis of infectious diseases~\cite{bail75},
the two basic models are \emph{infect and die} and \emph{infect forever}. In
the infect and die model, a disease or message is communicated for only a
single round and then the peer no longer participates. In the infect forever
model the peer can continue to communicate the message forever. For the very
basic case in the IME model, i.e.  a single message broadcast, the infect
forever model is applicable.  However, for a stream of messages that are being
broadcast then the situation is a non-trivial combination of infect and die
(because of the finite buffer limitation) and infect forever (because peers can
continue to communicate a message until they receive a subsequent message).
Also, epidemic information dissemination usually assumes that subsequent
generations of infected peers are selected at random from the population. In
the IME model this is not true because a peer can only receive a message when
it is online. 

The most closely related work is that of Yao, Leonard et. al.~\cite{yao}.  They
model heterogeneous user churn and local resilience of unstructured P2P
networks. They also concede early that balancing model complexity and its
fidelity is required to make advances in this area. They examine both the
Poisson and Pareto distribution for user churn and provide a deep analysis on
this front. Their work focuses on how churn affects connectivity in the
network and we have separated this aspect from our work and concentrated on
message throughput.

Other closely related work concerns mobile and ad hoc networks, and sensor
networks, because these applications require robust communication techniques
and tend to have limited buffer space at each node. The recent work of
Lindemann and Waldhorst~\cite{DBLP:conf/sigmetrics/Lindemann05} considers the
use of epidemiology in mobile devices with finite buffers and they follow the
seven degrees of separation system~\cite{papa}. In particular they use models
for ``power conservation" where each mobile device is ON with probability
$p_{ON}$ and OFF with probability $p_{OFF}$. Their analytical model gives very
close predictions to their simulation results. In our work we describe these
states using arrival rate, $\lambda$, and departure rate, $\mu$, which allows
us to naturally relate this to a rate of message arrivals, $\alpha$. We focus
solely on these parameters so that we can show precisely how they affect
message coverage rate.

Other closely related work such as in~\cite{constand} looks at the rate of file
transmission in a file sharing system that is based on epidemics. The use of
epidemics for large scale communication is also reviewed in~\cite{werner}. The
probabilistic multicast technique in~\cite{citeulike:392472} attempts to
increase the probability that peers receive messages for which they are
interested and to decrease the probability that peers receive messages for
which they are not interested. Hence it introduces a notion of membership which
is not too different to being online/offline. Autonomous Gossiping presented
in~\cite{karl} provides further examples of using epidemics for selective
information dissemination. 

\subsection{Organization}

In Section \ref{model} we describe our IME model and show the derivation of
equations that accurately predict its behavior. We compare the analytical
results with simulations. In Section \ref{coveragesec} we examine the use of
the model to choose message rates appropriate for the churn. In Section
\ref{kbuffersec} we provide a derivation of the $k$ buffer and multi-source
cases. We conclude the paper in Section \ref{conclusion} with some overall
observations and future work.

Table \ref{notation} provides the notation used in this paper. Generally, when
a function, $f$, is provided with a subscript $f_0$ or $f_1$ then the function
is representing either the offline or online case resp, e.g. $n_0$ and $n_1$ in
Table \ref{notation}. Later in the paper we extend this subscript notation to
represent more complicated cases. For stochastic processes like $X(t)$ we use
$\widehat{X}(t)$ as the expected value and we use $\bar{X}(t)$ as the
normalized expected value.
\begin{table}[htbp]
\caption{Notation}
\label{notation}
\begin{center}
\begin{tabular}{cp{0.8\linewidth}}\hline\hline
$N$ & number of peers \\
$\lambda$ & arrival rate of a peer \\
$\mu$ & departure rate of a peer \\
$n_0$ & mean number of peers offline \\
$n_1$ & mean number of peers online \\ 
$t$ & time \\
$X(t)$ & coverage of a single message at time $t$ \\ 
$\alpha$ & rate of message arrivals \\
$\widehat{C}$ & average coverage of a message in a message stream \\ 
$\widehat{C}_{base}$ & average base coverage a message in a message stream\\
$\widehat{C}^*$ & (extended) coverage rate \\
$\xi_T$ & fraction of messages of type $T$ (e.g. type $L1$, $1$, $0$, etc.)\\ 
$k$ & size of the buffer \\
$N_s$ & number of source peers\\
\hline\hline
\end{tabular}
\end{center}
\end{table}

\section{IME model and analytical formulation}
\label{model}

We begin this section with a basic description of the IME model using queueing
theory. We then provide a basic analysis for a single message broadcast.
Results from the basic analysis are used throughout the paper.  Simulation
results are compared to for each of the results. 

\subsection{Instantaneous message exchange model}

\begin{figure}[h]
\psfrag{l}{$\lambda$}
\psfrag{m}{$\mu$}
\psfrag{a}{$\tfrac{\lambda}{\lambda+\mu}N$}
\psfrag{b}{$\tfrac{\mu}{\lambda+\mu}N$}
\centerline{\includegraphics[width=0.5\linewidth]{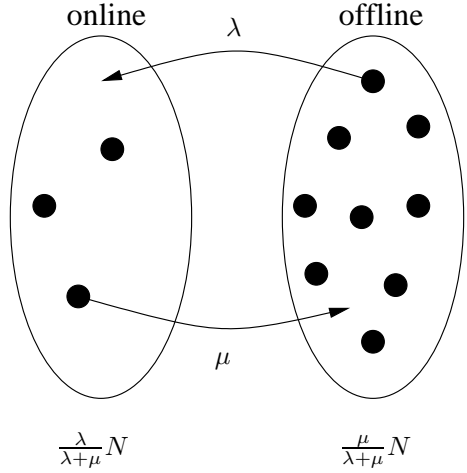}}
\caption{Peers are either online or offline; each peer independently arrives and
departs with rates $\lambda$ and $\mu$ respectively.}
\label{churn}
\end{figure}

\begin{figure}[h]
\scriptsize
\psfrag{Nl}{$N\,\lambda$}
\psfrag{(N-1)l}{$(N-1)\,\lambda$}
\psfrag{(N-2)l}{$(N-2)\,\lambda$}
\psfrag{2l}{$2\,\lambda$}
\psfrag{l}{$\lambda$}
\psfrag{Nm}{$N\,\mu$}
\psfrag{(N-1)m}{$(N-1)\,\mu$}
\psfrag{(N-2)m}{$(N-2)\,\mu$}
\psfrag{2m}{$2\,\mu$}
\psfrag{m}{$\mu$}
\psfrag{0}{$0$}
\psfrag{1}{$1$}
\psfrag{2}{$2$}
\psfrag{N-2}{$N-2$}
\psfrag{N-1}{$N-1$}
\psfrag{N}{$N$}
\includegraphics[width=\linewidth]{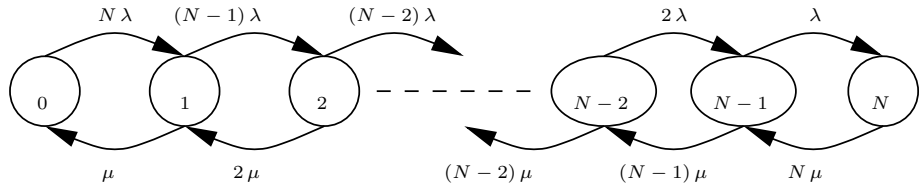}
\caption{State-transition diagram for $M/M/c/c/c$ queueing system (where $c=N$) queueing system.}
\label{queue}
\end{figure}

Consider a set of $N$ peers where each peer, $i\in\{1,\dotsc, N\}$ has a state,
$s_i\in\{0,1\}$ where 0 means offline and 1 means online. We say that a peer is
online or offline to mean which state it has. Let each peer change from offline
to online at a random time $t_0$ according to an ``arrival" rate, $\lambda$,
such that:
\[
\mathbb{P}[t_0 < t]=1-e^{-\lambda\,t},
\]
where $\mathbb{E}[t_0]=1/\lambda$ is the mean time
taken to change state from offline to online, with a 
cumulative distribution function given by the Poisson distribution\footnote{
In future work we shall investigate variations of the Pareto distribution for peer
lifetimes.}.  Similarly
$\mathbb{E}[t_1]=1/\mu$ is defined as the mean time to change state from online
to offline with
``departure" rate, $\mu$.  In other words, each peer spends a proportion,
$\tfrac{1}{\lambda} : \tfrac{1}{\mu}$, of its total time arriving and departing
respectively; shown in Figure \ref{churn}.
The peers are described by an $M/M/c/c/c$ queueing
system (where $c=N$) as shown in Figure \ref{queue}.

Using Figure \ref{queue}, if we let $n_1$ be the number of online peers, then we
can write:
\[
\begin{split}
p_n=\mathbb{P}[n_1=n]&=\binom{N}{n}\Big(\frac{\lambda}{\mu}\Big)^n p_0\\
p_0=\mathbb{P}[n_1=0]&=\left(\sum_{n=0}^N\binom{N}{n}\Big(\frac{\lambda}{\mu}\Big)^n\right)^{-1}\\
&=\Big(\frac{\lambda+\mu}{\mu}\Big)^{-N}.
\end{split}
\]
Hence
the system reaches an equilibrium when the number of online peers is
\[
n_1=\sum_{n=0}^{N} n\, p_n =  \tfrac{\lambda}{\lambda+\mu}N
\]
and equivalently when the number of offline peers is
\[
n_0=\tfrac{\mu}{\lambda+\mu}N=N-n_1.
\]

In this work we assume that the system is in equilibrium and we use $n_0$ and
$n_1$ as continuous variables.

\subsection{Single message broadcast}

Consider the case when a peer, called the source, is chosen uniformly at random
from $N$ and at time $t=0$ that peer generates a new message. 
The notion that messages can be transmitted instantly between online peers is
described by the rule: peer $i$ has received the message by time $t$ iff there
is some $t^\prime\leq t$ such that peer $i$ was online with some other online
peer that already had the message, at time $t^\prime$.  A peer that remains
offline up to (but not including) time $t$, cannot have received the message
before time $t$. 

Let $X(t)$ be a continuous time, discrete space, stochastic process that counts
the number of peers with the message by time $t$. We say that $X(t)$ is the
\emph{coverage} of the message at time $t$. At time $0$ the source peer is
initially offline with probability $n_0/N$ and online otherwise. Hence we
consider $\widehat{X}(t)$ which is the weighted average of two different
coverage functions, $X_0(t)$ and $X_1(t)$ for the initially offline and
initially online cases respectively.

For generality, we allow messages to be generated even if/while the peer is
offline. This is generally required in the case that, e.g., the peer is
generating or collecting data which is independent of whether the peer is
online or not. The special case when messages are not generated at a peer that
is offline, is then a simplification of the general case.
 
\subsubsection{Source peer starts online}

In this case, coverage starts at time $t=0$.  After time
$t_m=\frac{m}{n_0\,\lambda}$ (for some integer $m\geq 0$) the average number of
peers that have the message is 
\begin{equation}\label{xingrowth}
\begin{split}
\mathbb{E}\big[X_{1}(t_m)\big]=\widehat{X}_{1}(t_m)&=n_1+n_0(1-(1-1/n_0)^m) \\
\Rightarrow\widehat{X}_{1}(t)&=N - \frac{N\,\mu \,{\left( 1 - \frac{\lambda  + \mu }{N\,\mu } \right) }^{\frac{N\,t\,\lambda \,\mu }{\lambda  + \mu }}}{\lambda  + \mu } \\
\Rightarrow \bar{X}_1(t)= \lim_{N\rightarrow\infty}\frac{\widehat{X}_{1}(t)}{N}&=1 - \frac{e^{-t\,\lambda }\,\mu }{\lambda  + \mu }.
\end{split}
\end{equation}

In this work we mean field analysis in terms of $N$, since a P2P network is
expected to consist of a large number of nodes. The technique simplifies the
derivations in some cases; equations in terms of $N$ are however always
possible and we use both forms throughout.

\subsubsection{Source peer starts offline}

If the source peer starts
offline then it becomes online with probability $\lambda\, e^{-\lambda\, t}$ at
time $t$, i.e. at an average time of $1/\lambda$. Hence $\widehat{X}_{0}(t)$ is the
convolution with $\widehat{X}_{1}(t)$:
\begin{equation}\label{xoutgrowth}
\begin{split}
\widehat{X}_{0}(t)&=\int_0^t \widehat{X}_{1}(t-\tau)\,\lambda\, e^{-\lambda\, \tau}d\tau\\
&\quad +1-\int_0^t \lambda\, e^{-\lambda\, \tau}d\tau \\
\Rightarrow\bar{X}_0(t)=\lim_{N\rightarrow\infty}\frac{\widehat{X}_{0}(t)}{N}&=1 - \frac{e^{-t\,\lambda }\,\left(\lambda  + \mu  + t\,\lambda \,\mu\right) }{ \lambda  + \mu }
\end{split}
\end{equation}

The constant $1$ and last term of the integration represent the diminishing
constant which accounts for the ``fraction of the source peer" that has not yet
become online. Before the source peer becomes online, or more specifically at
time $t=0$, $\widehat{X}_{0}(0)=1$.  As $t$ increases, the probability
increases that the source peer becomes online and so too does the average
coverage increase, where the first term accounts for the fraction of the source
peer that has become online. 

\subsubsection{Expected coverage}

The expected coverage is the weighted average of Eqs.  \ref{xingrowth} and
\ref{xoutgrowth}:
\begin{equation}\label{xgrowth}
\begin{split}
\bar{X}(t)&=\tfrac{\lambda}{\lambda+\mu}\,\bar{X}_{1}(t)+\tfrac{\mu}{\lambda+\mu}\,\bar{X}_{0}(t)\\
&=1 - \frac{e^{-t\,\lambda }\,\mu \,\left( \mu  + \lambda \,\left( 2 + t\,\mu  \right)  \right) }{{\left( \lambda  + \mu  \right) }^2}.
\end{split}
\end{equation}

Of course, the observed coverage either starts from time $t=0$ or starts at an
average time of $1/\lambda$.  The expected coverage in Eq. \ref{xgrowth}
represents the average of these two cases. Figure \ref{chart1} shows 10
simulation runs when $\lambda=\mu=1$ and $N=1000$. The simulation tool is
described in the Appendix for reference. The points in a series indicate times when
a new peer received the message. The solid lines are $\widehat{X}_{1}(t)$ and
$\widehat{X}(t)$ (the functions are plotted starting from time 0.1). Note that
$\widehat{X}(t)$ is only an average, and is not representative of either the
online or offline cases.

\begin{figure}[htbp]
\psfrag{Xin}{$\widehat{X}_{1}(t)$}
\psfrag{Xout}{$\widehat{X}(t)$}
\psfrag{t}{$t$}
\includegraphics[width=\linewidth]{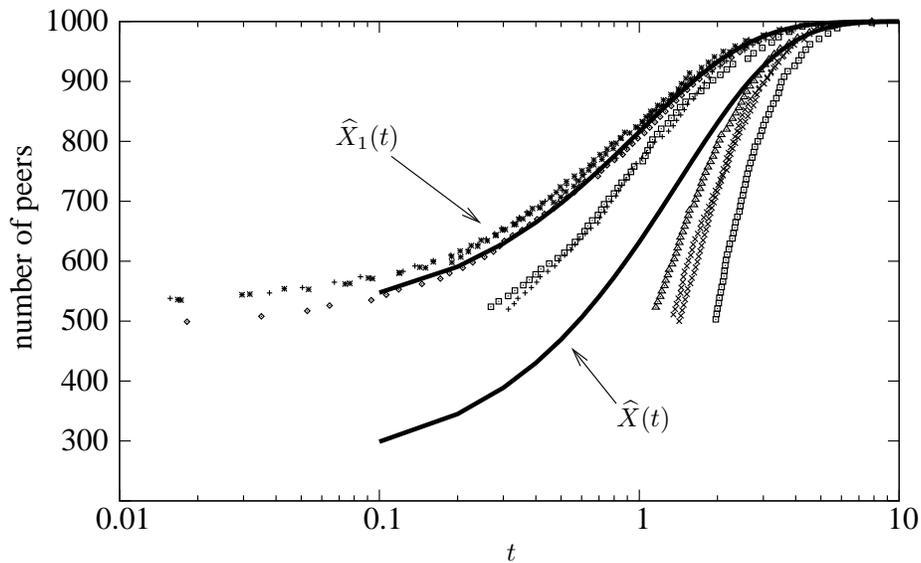}
\caption{Simulation runs (points) ($\lambda=\mu=1$, $N=1000$) and theoretical
curves (solid lines) for $\widehat{X}_{1}(t)$ and $\widehat{X}(t)$.}
\label{chart1}
\end{figure}

\subsubsection{Basic simulation results}

\subsection{Multiple message broadcast - unit message buffer}

\begin{figure}[h]
\psfrag{1}{$1$}
\psfrag{2}{$2$}
\psfrag{3}{$3$}
\psfrag{4}{$4$}
\psfrag{5}{$5$}
\psfrag{6}{$6$}
\psfrag{7}{$7$}
\psfrag{8}{$8$}
\psfrag{9}{$9$}
\psfrag{10}{$10$}
\psfrag{1/a}{$\frac{1}{\alpha}$}
\psfrag{1/m}{$\frac{1}{\mu}$}
\psfrag{1/l}{$\frac{1}{\lambda}$}
\psfrag{xout}{$\substack{\underbrace{\quad}\\\xi_{0}}$}
\psfrag{xin}{$\substack{\underbrace{\quad}\\\xi_{1}}$}
\includegraphics[width=\linewidth]{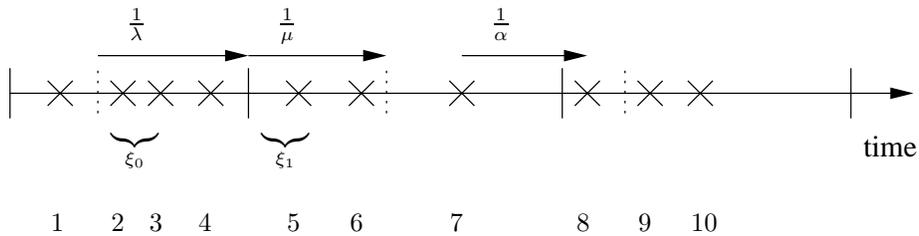}
\caption{Example realization in time of the source peer that changes between online
and offline. Arrival of a message at the source peer is depicted by a
$\times$.}
\label{messages}
\end{figure}

Let messages $\{1,2,\dotsc\}$ be generated at a source peer with rate $\alpha$
and consider a sequence of generation times $\{m_1,m_2,\dotsc\}$. Apply the
rule that each peer discards a current message in favor of a newer message and
does not receive messages that are older than the current message. A message is
said to be skipped by a peer if that message is not received by the peer
because a newer message has already been received.  Figure \ref{messages} shows
an example realization in time of events that occur at the source peer. It also
shows a number of numerical quantities that are important for the equations. A
solid vertical line represents the source peer moving from offline to online. 
A dashed vertical line represents moving from online to offline. Arrival of a new
message is shown by a $\times$ and the message numbers are given at the bottom
of the figure for reference.

We are interested in computing the average coverage of a message in a message
stream such as that shown in Figure \ref{messages}. Since messages are
discarded in favor of new ones, the coverage $X(t)$ of any given message is
limited. If $C_i(\alpha; \lambda,\mu)$ is the coverage of message $i$ in the
message stream of $M$ messages then we define 
\[
\widehat{C}=\lim_{M\rightarrow\infty}\frac{1}{M}\sum_{i=1}^M C_i
\]
as the average coverage of a message in a message stream.
 
In Figure \ref{messages} note that message $1$
arrives while the source peer is online, therefore it is immediately
received by all other peers that are online and the coverage of that
message at time $m_1$ (its arrival time) is immediately $n_1$. Shortly after
time $m_1$, the source peer goes offline. Message 1 continues to be
transmitted between new peers that enter the network and its coverage
increases. The arrival of messages 2, 3 and 4 does not hinder the transmission
of message 1 because the source peer is still offline. The coverage
of these messages is exactly 1.  However, shortly after the arrival of message
4, the source peer goes online. At this point, all online peers
receive message 4 and the coverage of message 4 jumps to $n_1$. The coverage of
message 4 continues to grow until the arrival of message 5, which is
immediately transmitted to all online peers (overwriting message 4 due to the unit
buffer restriction).  Message 5's coverage increases until the arrival of
message 6, and so on.

Note that messages 1, 6 and 8 are the last messages to arrive in each interval
for which the source peer is online. These messages continue to increase their
coverage until the source peer moves back online \emph{and} has a newly arrived
message in the mean time. E.g., if message 7 did not arrive then message 6
would have continued to grow until message 8 arrived.  Also note that messages
2, 3 and 9 were not received by any other peers and never will be, their
coverage remains at 1.
 
We identify four categories of messages (the naming scheme simplifies our
notation later), listed in Table \ref{messagecat1}.
\begin{table}[htbp]
\caption{Message categories}
\label{messagecat1}
\begin{center}
\begin{tabular}{cp{0.8\linewidth}}\hline\hline
$L1$ & A message that is the last one to arrive before
the peer moves from online to offline (e.g. messages 1, 6 and 8).\\
$1$ & Messages that arrive while the peer is online but not
a $L1$ message (e.g. message 5).\\ 
$L0$ & A message that is the last one to arrive before
the peer moves from offline to online (e.g. messages 4, 7 and 10).\\
$0$ & Messages that arrive while the peer is offline but not
a $L0$ message (e.g. messages 2, 3 and 9).\\
\hline\hline
\end{tabular}
\end{center}
\end{table}

When only one message arrives in an interval (by interval we mean either the
online or offline time interval) then that message is a $L1$ or $L0$
(i.e., in this case there are no $1$ or $0$ messages in that interval).

Intuitively,
\[
\lim_{\alpha\rightarrow 0} \widehat{C}(\alpha\,;\,\lambda,\mu) = N
\]
because at low message arrival rates each message has ample time to cover all
peers. Of course, when $\alpha\rightarrow 0$ then the message throughput is low
which is undesirable.

As a consequence of our IME model, note that any message that achieves a
coverage of greater than 1 will achieve a coverage of at least $n_1$. We call
this the \emph{base coverage} and we note that the average base coverage is the
average coverage of a message as the message rate becomes large:
\[
\widehat{C}_{base}(\lambda,\mu)=\lim_{\alpha\rightarrow \infty} \widehat{C}(\alpha\,;\,\lambda,\mu) = 
\frac{\lambda}{\lambda+\mu}\,n_1+\frac{\mu}{\lambda+\mu}.
\]
All $1$ messages reach exactly $n_1$ peers (immediately on their arrival), all
$0$ messages reach only 1 peer (the source peer) and the number of $L1$ and
$L0$ messages becomes negligible.

Since in the IME model, $\widehat{C}$ does not approach 0 as
$\alpha\rightarrow\infty$, the \emph{natural coverage rate}
$\alpha,\widehat{C}$ (or message-peer throughput) is unbounded. However, we observe that in this case the
average coverage approaches the constant $\widehat{C}_{base}$ and so while the
rate of messages is arbitrarily large, the messages are received by only a
fixed fraction of the peers.

For these reasons we are interested in coverage achieved \emph{beyond} the base
coverage and we formulate what we call the mean \emph{extended coverage rate}:
\begin{equation}\label{optproblem}
\widehat{C}^*(\alpha\,;\,\lambda,\mu)=\alpha\,\frac{\widehat{C}(\alpha\,;\,\lambda,\mu)-\widehat{C}_{base}(\lambda,\mu)}
{N-\widehat{C}_{base}(\lambda,\mu)}.
\end{equation}
In this paper we simply refer to Eq. \ref{optproblem} simply as the coverage
rate which has units message-peers per time unit. When $\alpha$ is small, while
a single message will cover all of the peers, there are not many messages and
the overall number of messages received by the peers is small; ultimately the
coverage rate falls to zero.  When $\alpha$ is large, while a number of peers
$n_1$, at any one time, may receive a large number of messages the actual
coverage of a single message is at most $\widehat{C}_{base}$ and the extended
coverage drops to 0.

To analyze Eq. \ref{optproblem}, we derive an equation for
$\widehat{C}(\alpha\,;\,\lambda,\mu)$ by combining individual equations for the
different message categories.

\subsubsection{Fraction of appearance for each message category}
\label{mescateg}

Clearly, $\frac{\mu}{\lambda+\mu}$ of all messages will arrive while the source
peer is offline, i.e. they are $0$ and $L0$ messages, and similarly for $1$ and
$L1$ messages. We need to know the fraction for each category; in Figure
\ref{messages} we use $\xi_{0}$ to represent the fraction of messages that
are $0$ and $\xi_{1}$ to represent the fraction of messages that are $1$. Then
the fraction that are $L0$ is:
\begin{equation}\label{last-out-pre}
\xi_{L0}=\frac{\mu}{\lambda+\mu}-\xi_{0}
\end{equation}
and similarly for the $L1$ fraction:
\begin{equation}\label{last-in-pre}
\xi_{L1}=\frac{\lambda}{\lambda+\mu}-\xi_{1}.
\end{equation}

For a given message rate $\alpha$, in the time interval $1/\lambda$ we have
$\alpha/\lambda$ messages arriving on average. The fraction $\xi_{0}$ is
derived by summing the individual probabilities of $k\ge 2$ messages arriving
before the state change from offline to online occurs, where we are weighting
the event that $k-1$ messages become $0$ messages. This is divided by the
average number of messages that arrive because we are interested in the
fraction of such messages, not the total. The equation becomes:
\begin{equation}\label{out}
\begin{split}
\xi_{0}&=\tfrac{\mu }{\lambda  + \mu }\,\tfrac{\lambda }{\alpha }\,
   \int _{0}^{\infty } \sum_{j = 2}^{\infty }\left( j - 1 \right) \,e^{-\alpha \,t}\,
           \frac{{\left( \alpha \,t \right) }^{j}}{j!} \, \lambda \,e^{-\lambda \,t}  \,
      dt\\
&=\tfrac{\mu }{\lambda  + \mu }\,\tfrac{\lambda }{\alpha }\,
   \int _{0}^{\infty }
(-1 + e^{- t\,\alpha } + t\,\alpha) \lambda \,e^{-\lambda \,t}
\,dt\\
&=\frac{\alpha \,\mu }{\left( \alpha  + \lambda  \right) \,\left( \lambda  + \mu  \right) }.
\end{split}
\end{equation}
Similarly:
\begin{equation}\label{in}
\begin{split}
\xi_{1}&=\tfrac{\lambda }{\lambda  + \mu }\,\tfrac{\mu }{\alpha }\,
   \int _{0}^{\infty } \sum_{j = 2}^{\infty }\left( j - 1 \right) \,e^{-\alpha \,t}\,
           \frac{{\left( \alpha \,t \right) }^{j}}{j!} \, \mu \,e^{-\mu \,t}  \,
      dt\\
&=\frac{\alpha \,\lambda }{\left( \alpha  + \mu  \right) \,\left( \lambda  + \mu  \right) }.
\end{split}
\end{equation}

Hence Eqs. \ref{last-out-pre} and \ref{last-in-pre} become:
\begin{equation}\label{last-out}
\xi_{L0}=\frac{\lambda \,\mu }{\left( \alpha  + \lambda  \right) \,\left( \lambda  + \mu  \right) }
\end{equation}
and
\begin{equation}\label{last-in}
\xi_{L1}=\frac{\lambda \,\mu }{\left( \alpha  + \mu  \right) \,\left( \lambda  + \mu  \right) }
\end{equation}

\subsubsection{Coverage of each message category}

The average coverage of $0$ messages, $\widehat{C}_{0}$, is trivially 1 since
these messages do not have a chance to be communicated to other peers.  The
average coverage of $1$ messages, $\widehat{C}_{1}$, follows the single message
coverage function from Eq. \ref{xingrowth}, $\widehat{X}_1(t)$, where the
average time is $1/\alpha$.  Since $\widehat{X}_1(t)$ is non-linear we
integrate the growth over all possible times:
\[
\begin{split}
\widehat{C}_{1}&=\int _{0}^{\infty }\widehat{X}_{1}(t)\,\alpha \,e^{-\alpha \,t}\, dt\\
&=N + \tfrac{N\,\alpha \,\mu }{-\left( \alpha \,\left( \lambda  + \mu  \right)  \right)  + 
     N\,\lambda \,\mu \,\log (-\left( \frac{\lambda  + \mu  - N\,\mu }{N\,\mu } \right) )}\\
\Rightarrow \bar{C}_1 = \lim_{N\rightarrow \infty}\frac{\widehat{C}_{1}}{N}&=\frac{\lambda \,\left( \alpha  + \lambda  + \mu  \right) }
   {\left( \alpha  + \lambda  \right) \,\left( \lambda  + \mu  \right) }
\end{split}
\]
The coverage of $L1$ and $L0$ messages is more difficult to model because their
average coverage time is affected by whether the peer is online or offline when
the subsequent message arrives. If the peer is offline then the time increases
by an amount given by the average time for the peer to become online again. We
therefore further categorize these messages as $L1-1$, $L1-0$, $L0-1$ and
$L0-0$ messages, depending on whether the peer is online or offline when the
subsequent message arrives. 

\begin{figure}[htbp]
\psfrag{a}{$\alpha$}
\psfrag{l}{$\frac{1}{\lambda}$}
\psfrag{l+m}{$\frac{1}{\lambda}+\frac{1}{\mu}$}
\psfrag{2l+m}{$\frac{2}{\lambda}+\frac{1}{\mu}$}
\psfrag{2(l+m)}{$\frac{2}{\lambda}+\frac{2}{\mu}$}
\psfrag{t}{$t$}
\includegraphics[width=\linewidth]{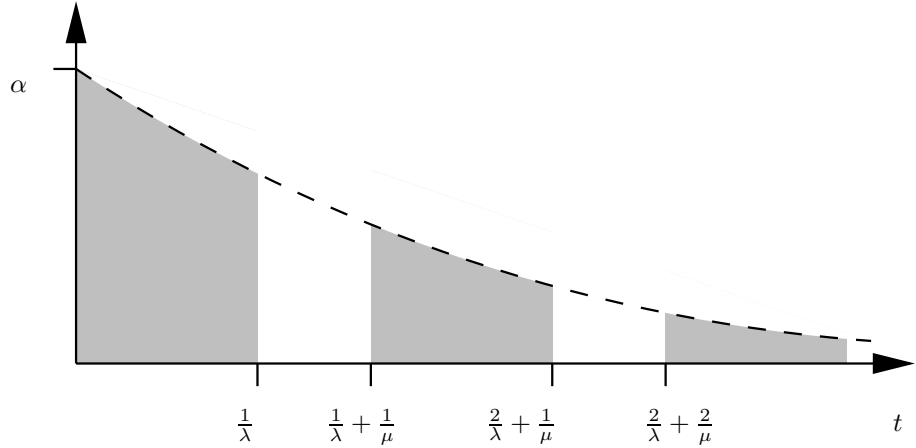}
\caption{Probability distribution, $\alpha e^{-\alpha t}$. Shaded areas
represent average times when the peer is offline.} 
\label{exp}
\end{figure}

For $L1-0$ messages we integrate over the shaded regions shown in Figure
\ref{exp} and we essentially round up to the $1/\lambda$ interval. For $L1-1$
messages we integrate exactly to the time $t$. The integration is similar for
$L0-0$ and $L0-1$ messages, except that the integration time $t=0$ begins at
the beginning of a $1/\mu$ interval rather than at the beginning of a
$1/\lambda$ interval.
Let $a=\frac{1}{\lambda}+\frac{1}{\mu}$, then the integrations become: 
\[
\begin{split}
\widehat{C}_{L1-0}&=\sum_{j = 0}^{\infty }\widehat{X}_{1}(j\,a + \frac{1}{\lambda })\,
     \int _{j\,a}^{j\,a + \frac{1}{\lambda }}\alpha \,e^{-\alpha \,t}\,dt,\\
\widehat{C}_{L1-1}&=\sum_{j = 0}^{\infty }\int _{j\,a + \frac{1}{\lambda }}^{\left( j + 1 \right) \,a}
      \widehat{X}_{1}(t)\,\alpha \,e^{-\alpha \,t}\,dt,\\
\widehat{C}_{L0-0}&=\sum_{j = 0}^{\infty }\widehat{X}_{1}(\left( j + 1 \right) \,a)\,
     \int _{j\,a + \frac{1}{\mu }}^{\left( j + 1 \right) \,a}\alpha \,e^{-\alpha \,t}\,dt,\\
\widehat{C}_{L0-1}&=\sum_{j = 0}^{\infty }\int _{j\,a}^{j\,a + \frac{1}{\mu }}
      \widehat{X}_{1}(t)\,\alpha \,e^{-\alpha \,t}\,dt.
\end{split}
\]
We show here only the expressions when $N\rightarrow \infty$. For
$L1$ messages:
\begin{multline}
\bar{C}_{L1}=\lim_{N\rightarrow\infty}\frac{\widehat{C}_{L1-0}+\widehat{C}_{L1-1}}{N}=\\1 + \frac{\left( \alpha  + e^{\frac{\alpha  + \lambda }{\mu }}\,
         \left( \lambda  - e^{\frac{\alpha }{\lambda }}\,\left( \alpha  + \lambda  \right)  \right)  \right) \,\mu }{
      \left( -1 + e^{\frac{\left( \alpha  + \lambda  \right) \,\left( \lambda  + \mu  \right) }{\lambda \,\mu }} \right) \,
      \left( \alpha  + \lambda  \right) \,\left( \lambda  + \mu  \right) }.
\end{multline}
Similarly for $L0$ messages:
\begin{multline}
\bar{C}_{L0}=\lim_{N\rightarrow\infty}\frac{\widehat{C}_{L0-0}+\widehat{C}_{L0-1}}{N}=\\
\frac{\frac{\left( e^{\frac{\alpha }{\lambda }}\,\left( \left( -1 + e \right) \,\alpha  - \lambda  \right)  + 
          \lambda  \right) \,\mu }{-1 + e^{\frac{\left( \alpha  + \lambda  \right) \,\left( \lambda  + \mu  \right) }{\lambda \,\mu }}}
      + \lambda \,\left( \alpha  + \lambda  + \mu  \right) }{\left( \alpha  + \lambda  \right) \,\left( \lambda  + \mu  \right) }
\end{multline}

\subsubsection{Total coverage equation}

Each of the coverage functions contribute according to the fractions 
in Eqs. \ref{in}, \ref{out}, \ref{last-in} and \ref{last-out}. Hence:
\begin{multline}\label{coverage}
\widehat{C}=\xi_{1}\,\widehat{C}_{1}+
\xi_{L1}\,\big(\widehat{C}_{L1-0}+\widehat{C}_{L1-1}\big)+\\
\xi_{0}+
\xi_{L0}\,\big(\widehat{C}_{L0-0}+\widehat{C}_{L0-1}\big).
\end{multline}

Note that for $\bar{C}$ the fractional term $\xi_0$ becomes 0 and all other
terms are interchangeable.

\begin{figure}[htbp]
\psfrag{l=2.0}{$\lambda=2.0$}
\psfrag{l=1.0}{$\lambda=1.0$}
\psfrag{l=0.5}{$\lambda=0.5$}
\psfrag{alpha}{$\alpha$}
\includegraphics[width=\linewidth]{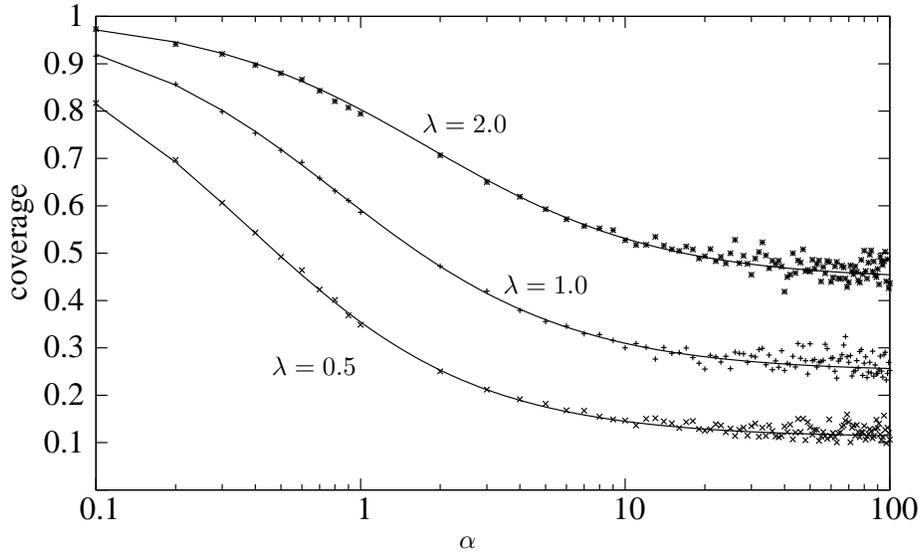}
\caption{Averaged simulation runs (points) and theoretical curves (solid lines)
for $\bar{C}$ with $\mu=1.0$ and $\lambda$ as shown.}
\label{chart2}
\end{figure}

While inspection of the final coverage equation does offer some insight, we
omit it due to its complex structure.  Figure \ref{chart2} shows the results
from simulations. Each point is the average of 10 trials with $N=100$,
$\mu=1.0$, 1000 messages transmitted and other parameters as shown. The
coverage is normalized.  Clearly, as $\alpha$ increases then the coverage
decreases. Note that as $\alpha\rightarrow\infty$, the coverage limits to
$\widehat{C}_{base}$. The precision of the simulation results decreases as
$\alpha$ increases because the simulation is run for a fixed number of messages
and hence an increased $\alpha$ leads to a decreased run time. The solid lines
represent the coverage as evaluated from Eq. \ref{coverage}.

Clearly, if a specific coverage is required (at least) by an application then
$\alpha$ takes a limited range. E.g., if the application requires a coverage of
at least 0.6 of the total peers, in a case where $\lambda=\mu=1.0$ then from
Figure \ref{chart2}, $\alpha$ is limited to be less than roughly 1.

\begin{figure*}[t!]
\begin{center}
\psfrag{y=a}{$y=\alpha$}
\psfrag{l=2.0}{$\lambda=2.0$}
\psfrag{l=1.0}{$\lambda=1.0$}
\psfrag{l=0.5}{$\lambda=0.5$}
\psfrag{alpha}{$\alpha$}
\subfigure[$\mu=1.0$]{\label{chart3}\includegraphics[width=0.45\linewidth]{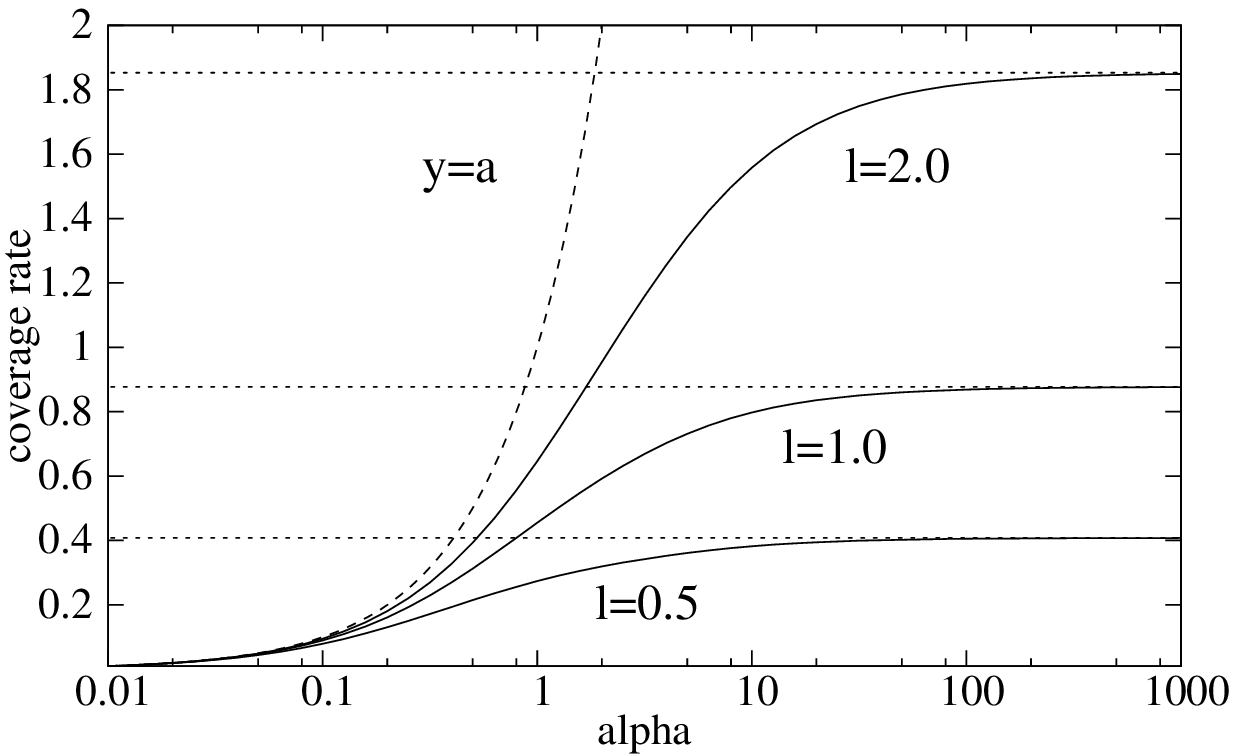}}
\subfigure[$\mu=100.0$]{\label{chart4}\includegraphics[width=0.46\linewidth]{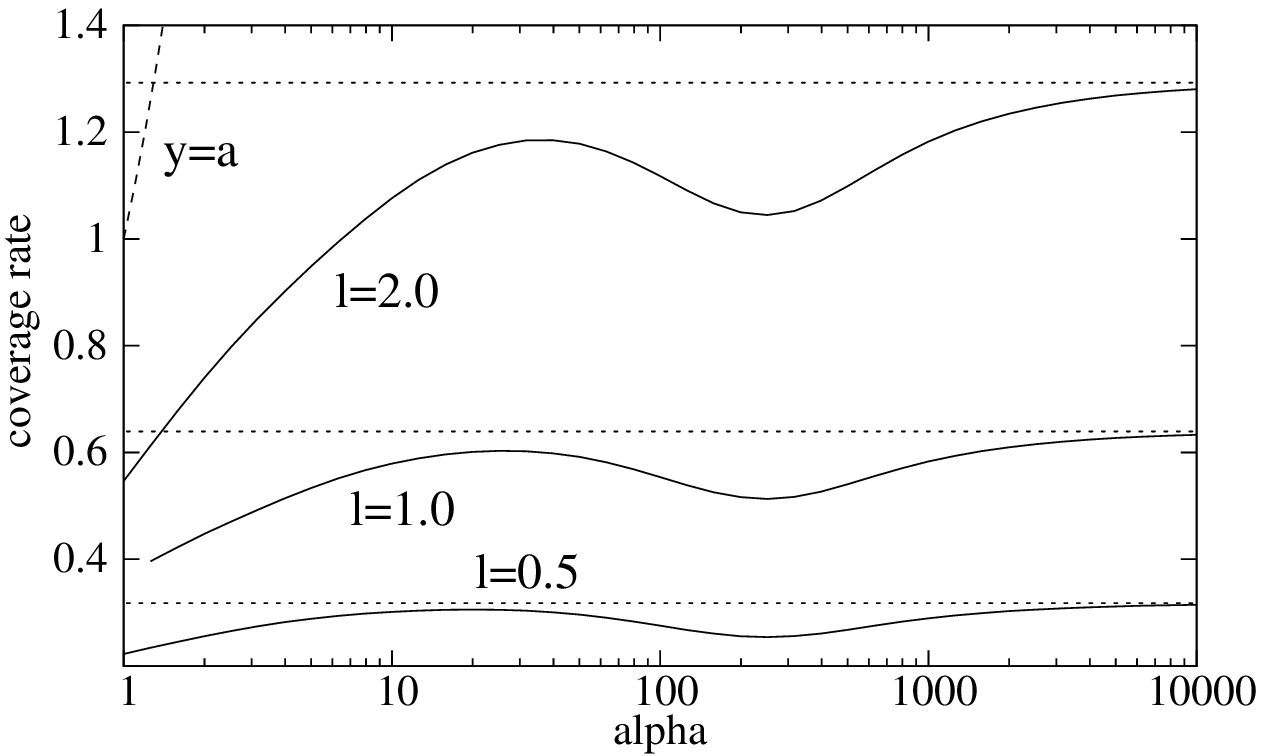}}
\caption{Coverage rate, $\widehat{C}^*$, (solid lines) with $\lambda$ as shown.}
\end{center}
\end{figure*}

\subsubsection{Coverage rate}

\label{coveragesec}

We plot the coverage rate from Eq. \ref{optproblem} in Figure \ref{chart3}.
When $\alpha$ is low, the coverage rate is close to $\alpha$. The coverage rate
is never equal to $\alpha>0$ and Figure \ref{chart3} shows $y=\alpha$ as a
reference. As $\lambda$ increases (or equivalently as $\mu$ decreases) then it
is possible to achieve a higher coverage rate because peers spend more of their
time in the network. 

The coverage rate saturates with large $\alpha$ and we have:
\begin{equation}\label{limit}
\lim_{\alpha\rightarrow\infty}\widehat{C}^*(\alpha;\lambda,\mu)=
\frac{\lambda \,\left( -\mu  + e\,\left( 2\,\lambda  + \mu  \right)  \right) }
   {e\,\left( 2\,\lambda  + \mu  \right) }.
\end{equation}
These limits are shown in Figures \ref{chart3} and \ref{chart4} as horizontal
dashed lines.

Note that: 
\[
\lim_{\alpha,\mu\rightarrow\infty}\widehat{C}^*(\alpha;\lambda,\mu)=\frac{\left( -1 + e \right) \,\lambda }{e},
\]
which is a consequence of there being no message transmission rate limits on
individual peers.

Figure \ref{chart4} shows that the coverage rate exhibits a local maximum which
approaches Eq. \ref{limit}. Hence, for parameters in these ranges it is possible
to achieve a close to maximum coverage rate at a relatively small $\alpha$.
For example, when $\lambda=0.5$ and $\mu=100.0$ then from Figure \ref{chart4}
we have $\alpha\approx 20$ to achieve a coverage rate that is very close to the
maximum, which would not otherwise be met until $\alpha>>1000$.

Simulation results (not reported here) show that these coverage rates are
highly susceptible to deviations from the average coverage. Therefore, this
analysis can serve as a rough guide, in the sense that while a value of
$\alpha$ may be computed as giving a particular coverage rate, an observed
coverage rate (which is necessarily over a finite range) is likely to deviate
from the theoretical prediction.

\section{Message buffer and multiple sources}

In this section we formulate the coverage for the cases when peers can buffer
$k$ messages and when there are multiple sources of messages.

\subsection{Using a $k$-buffer}
\label{kbuffersec}

We calculate coverage for the $k$-buffer case similarly to the unit buffer
case, dividing messages into separate categories. In this section we redefine
the $\xi_{in}$ and $\xi_{out}$ fractions to be with respect to the $k$-buffer
case. We also define further fractions.

\subsubsection{Message categories and their fractions}

As in the unit buffer case we consider the fraction of messages that arrived in
the period when the source peer was either online or offline and was pushed
from the buffer before the peer changed its state; which we call $\xi_{1-k}$
and $\xi_{0-k}$, meaning fraction of $1$ and $0$ messages respectively, where
buffer size is equal to $k$. In this case we calculate the fraction $\xi_{0-k}$
by summing the individual probabilities of $s\ge k$ messages arriving before
the state changes from offline to online, where we are weighting the event that
$j-k$ messages become $0$ messages. As in the unit buffer case, this is divided
by the average number of messages that arrive. So, the equations become:   

\begin{equation}\label{out_k}
\small
\begin{split}
\xi_{0-k}&=\tfrac{\mu }{\lambda  + \mu }\,\tfrac{\lambda }{\alpha }\,
   \int _{0}^{\infty } \sum_{j = (k+1)}^{\infty }\left( j - k \right) \,e^{-\alpha \,t}\,
           \frac{{\left( \alpha \,t \right) }^{j}}{j!} \, \lambda \,e^{-\lambda \,t}  \,
      dt.\\
\end{split}
\end{equation}

\noindent Similarly:

\begin{equation}\label{in_k}
\small
\begin{split}
\xi_{1-k}&=\tfrac{\lambda }{\lambda  + \mu }\,\tfrac{\mu }{\alpha }\,
   \int _{0}^{\infty } \sum_{j = (k+1)}^{\infty }\left( j - k \right) \,e^{-\alpha \,t}\,
           \frac{{\left( \alpha \,t \right) }^{j}}{j!} \, \mu \,e^{-\mu \,t}  \,
      dt.\\
\end{split}
\end{equation}

Unlike the unit buffer case we now have to consider a number $k$ of both $L1$
and $L0$ messages. We say that a message is an $L1-i$ or $L0-i$ message to mean
that $(i-1)$ messages arrived before the peer changes the state. 

Each $L1-i$ and $L0-i$ message will have it's own rate and coverage because
each of them will have a different coverage time. This is because there is,
e.g., an average time of $1/\alpha$ between message $L1-k$ and message
$L1-(k-1)$, and so on. 

\begin{figure}[h]
\psfrag{1}{$1$}
\psfrag{2}{$2$}
\psfrag{3}{$3$}
\psfrag{4}{$4$}
\psfrag{xin-2}{$\substack{\underbrace{\quad\quad\quad\quad}\\\xi_{1-2}}$}
\psfrag{xlast-in-1}{$\substack{\underbrace{\quad}\\\xi_{L1-1}}$}
\psfrag{xlast-in-2}{$\substack{\underbrace{\quad}\\\xi_{L1-2}}$}
\centerline{\includegraphics[width=0.7\linewidth]{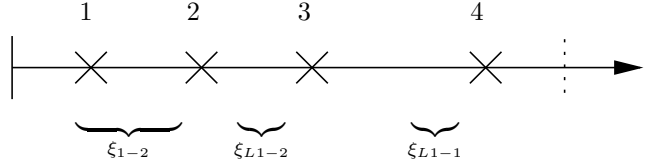}}
\caption{Arrival of messages, while the source peer is online, with $k=2$
and showing the fractions of different message types.}
\label{k2message}
\end{figure}

An example for $k=2$ is shown
in Figure \ref{k2message}. In the figure:
\begin{description}
\item[$1-2$]:  Message 1 is propagated until messages 3 comes,
similarly message 2 propagates until message 4 is generated. In other words
both of messages 1 and 2 get coverage until 2 more messages arrive.
\item[$L1-2$]: Message 3 is propagated until the peer goes offline, and then it
continues to propagate until at least one more message has arrived after
message 4 and the peer has come back online.
\item[$L1-1$]: Message 4 is propagated until at least two more
messages are generated, similarly to the previous example.
\end{description}

\begin{figure}[h]
\psfrag{k}{$k$}
\psfrag{i}{$i$}
\psfrag{i-1}{$i-1$}
\psfrag{1}{$1$}
\psfrag{xin-i}{$\substack{\underbrace{\quad\quad\quad\quad\quad\quad\quad\quad}\\\xi_{1-i}}$}
\psfrag{xin-i-1}{$\substack{\underbrace{\quad\quad\quad\quad\quad\quad\quad\quad\quad\quad}\\\xi_{1-(i-1)}}$}
\centerline{\includegraphics[width=0.7\linewidth]{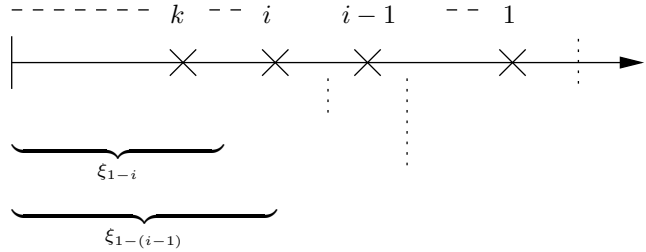}}
\caption{Computing the fraction $\xi_{L1-i}$.}
\label{kimessage}
\end{figure}

We derive $\xi_{L1-i}$, $i=1,2,\dotsc,k$, as shown on Figure \ref{kimessage}:

\begin{equation}\nonumber
\xi_{L1-i}=\xi_{1-(i-1)} -\xi_{1-i}.
\end{equation}
Similarly:
\begin{equation}\nonumber
\xi_{L0-i}=\xi_{0-(i-1)} -\xi_{0-i}.
\end{equation}

In the above equations $\xi_{1-(i-1)}$, $\xi_{1-i}$ and $\xi_{0-(i-1)}$,
$\xi_{0-i}$ fractions  are calculated using equations \ref{in_k} and
\ref{out_k}, respectively, substituting for $k$.

\subsubsection{Coverage of $1$ messages}

All $1-k$ messages propagate until at least $k$ subsequent messages have been
generated. So we look at the probability of the $k$-th message arriving
at time $t$. We use the Erlang distribution to compute the probability of the
$k$-th message arriving: 

\begin{equation}
\label{erlang-cov}
\begin{split}
\widehat{C}_{1-k}&=\int _{0}^{\infty }\widehat{X}_{1}(t)\,\frac {\alpha^k t^{k-1}}{(k-1)!}\,e^{-\alpha \,t}\, dt.\\
\end{split}
\end{equation}

\subsubsection{Coverage of $L1-i$ messages}

As in the unit buffer case we consider $L1-1-i$ and $L1-0-i$ messages.
A message is propagated while it exists in the buffer; to be pushed out of the
buffer, $k$ more messages have to arrive into the buffer. Because we are
considering only time periods after the period in which message $i$ was
generated, we know that $(i-1)$ messages have already arrived. Thus only $k-(i-1)$
more messages have to arrive to push the $L1-i$ message from the buffer. We
are interested in the probability of the $i$-th message arriving in a
subsequent online period. The propagation starts from the point when the $i$-th
message was generated, i.e we are interested in the coverage after time $t +
\frac{i-1}{\alpha}$. We use the Erlang distribution again:

\begin{multline}\nonumber
\small
\widehat{C}_{L1-1-i}=
\sum_{j = 0}^{\infty }\int _{j\,a + \frac{1}{\lambda }}^{\left( j + 1 \right) \,a}
      \widehat{X}(t+\,\frac{i-1}{\alpha})\,\frac{\alpha^{k-(i-1)} t^{k-i}}{(k-i)!}\,e^{-\alpha \,t}\,dt.\\
\end{multline}

Similarly for $L1-0-i$, we modify the result for $L1-0$ to get:

\begin{multline}\nonumber
\small
\widehat{C}_{L1-0-i}= \\
\sum_{j = 0}^{\infty }\widehat{X}_{1}(j\,a + \frac{1}{\lambda } + \frac{i-1}{\alpha})\,
     \int _{j\,a}^{j\,a + \frac{1}{\lambda }}\,\frac{\alpha^{k-(i-1)} t^{k-i}}{(k-i)!}\,e^{-\alpha \,t}\,dt.
\end{multline}

\subsubsection{Coverage of $L0-i$ messages}

We use equations derived for unit buffer case, except as for $L1-i$ messages
we wait until the $(k-(i-1))$-th message arrives:

\begin{multline}\nonumber
\widehat{C}_{L0-0-i}=
\sum_{j = 0}^{\infty }\widehat{X}(\left( j + 1 \right) \,a)\,
     \int _{j\,a + \frac{1}{\mu }}^{\left( j + 1 \right) \,a}\,\frac{\alpha^{k-(i-1)} t^{k-i}}{(k-i)!}\,e^{-\alpha \,t}\,dt,
\end{multline}

\begin{multline}\nonumber
\widehat{C}_{L0-1-i}=
\sum_{j = 0}^{\infty }\int _{j\,a}^{j\,a + \frac{1}{\mu }}
      \widehat{X}(t)\,\frac{\alpha^{k-(i-1)} t^{k-i}}{(k-i)!}\,e^{-\alpha \,t}\,dt.
\end{multline}

Note that the time within $\widehat{X}()$ has no offset because it does not
matter how recent a message arrived before the peer became online,
propagation starts only after the peer becomes online.

\subsubsection{Total coverage equation for $k$-buffer case}

We add the coverage of each message type to get total coverage:
\begin{multline}\nonumber
\widehat{C}(\alpha;\lambda,\mu, k)=\xi_{1-k}\,\widehat{C}_{1-k}+\\
\sum_{j = 1}^{k }\xi_{L1-j}\,\big(\widehat{C}_{L1-0-j}+\widehat{C}_{L1-1-j}\big)+\\
\sum_{j = 1}^{k }\xi_{L0-j}\,\big(\widehat{C}_{L0-0-j}+\widehat{C}_{L0-1-j}\big).
\end{multline}

\begin{figure}[h]
\psfrag{k=1}{$k=1$}
\psfrag{k=2}{$k=2$}
\psfrag{k=3}{$k=3$}
\psfrag{k=5}{$k=5$}
\psfrag{a}{$\alpha$}
\includegraphics[width=\linewidth]{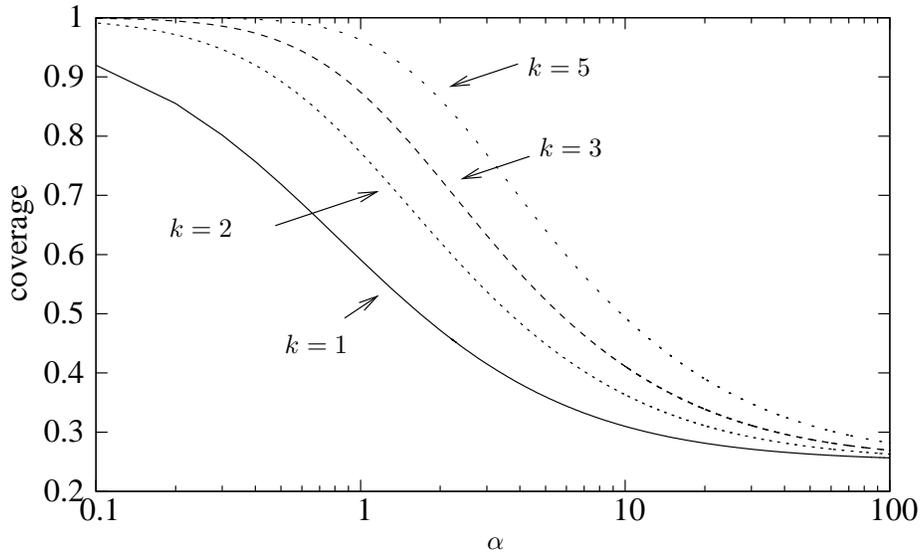}
\caption{Theoretical results for $\widehat{C}(\alpha;\lambda,\mu, k)$ with ${\mu=\lambda=1.0}$, $k$ as shown.}
\label{k_buf}
\end{figure}

\begin{figure}[h]
\psfrag{a}{$\alpha$}
\psfrag{l=0.5}{$\lambda=0.5$}
\psfrag{l=1.0}{$\lambda=1.0$}
\psfrag{l=2.0}{$\lambda=2.0$}
\psfrag{coverage}{coverage}
\includegraphics[width=\linewidth]{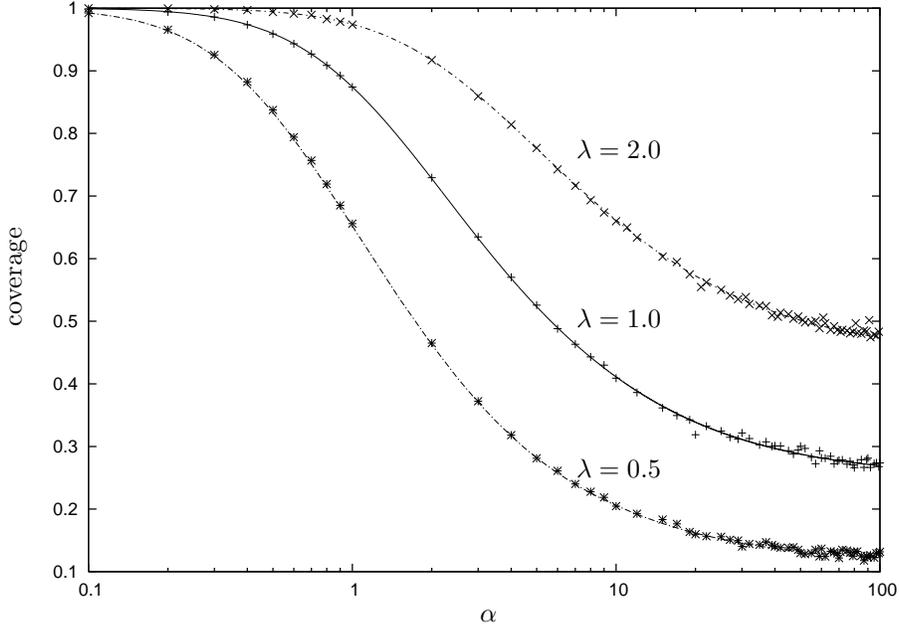}
\caption{Averaged simulation runs (points) and theoretical curves (solid lines)
for $\widehat{C}(\alpha;\lambda,\mu,k)$ with ${\mu=1.0}$, $k=3$ and $\lambda$ as shown.}
\label{3_buf}
\end{figure}

\begin{figure}[h]
\psfrag{l=0.5}{$\lambda=0.5$}
\psfrag{l=1.0}{$\lambda=1.0$}
\psfrag{l=2.0}{$\lambda=2.0$}
\psfrag{coverage}{coverage}
\psfrag{k}{$k$}
\includegraphics[width=\linewidth]{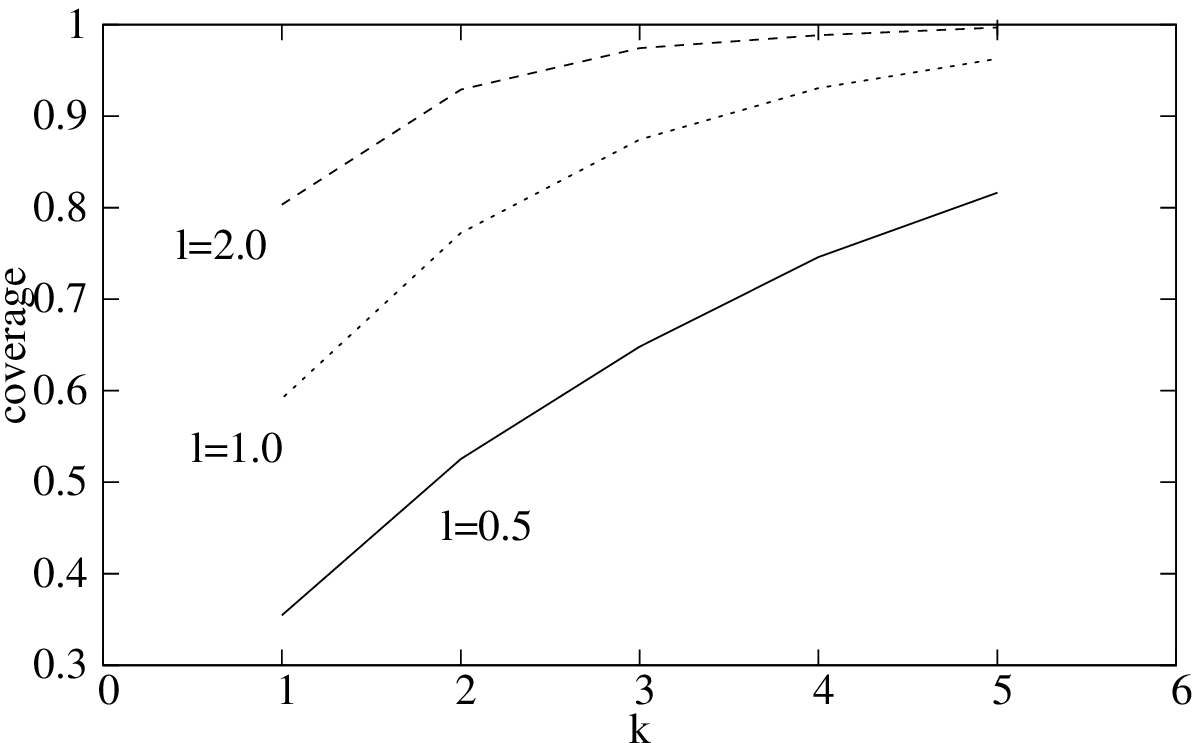}
\caption{Theoretical increase in $\widehat{C}(\alpha;\lambda,\mu,k)$ versus $k$
for $\mu=1.0$ and $\lambda$ as shown.}
\label{lambda}
\end{figure}

Figure \ref{k_buf} shows theoretical results for $k$ values 1,2,3 and 5 with
fixed ${\lambda=\mu=1}$.  Figure \ref{3_buf} shows a comparison of simulation
runs with $N=100$, ${\mu=1}$ and 1000 messages in the network.  Figure
\ref{lambda} shows the theoretical increase in coverage as $k$ increases, for
various $lambda$. E.g., to achieve a coverage of at least 0.7 when
$\lambda=0.5$ we need to have a buffer size of at least 4. Clearly, increasing
$k$ increases the coverage. Furthermore, $\lim_{\alpha\rightarrow
\infty}\widehat{C}(\alpha;\lambda,\mu,k)=\lim_{\alpha\rightarrow
\infty}\widehat{C}(\alpha;\lambda,\mu,1)$ since the fraction of $L1-k-i$ and
$L0-k-i$ messages becomes insignificant.

\subsection{Multiple source model}
\label{multsoursec}

In this section we consider the case when multiple peers are generating
messages. We maintain the message arrival rate $\alpha$ on a network wide
basis, i.e. if there are $N_s$ sources in the network each peer generates
messages with rate $\tfrac{\alpha}{N_s}$ rate.

We make the following simplifications:
\begin{itemize}
\item We ignore messages that occur on the peer
when the peer is offline. These kinds of messages were included in the previous
sections and could be removed if desired. This simplification limits our model
in this section to applications where messages are only generated on peers that
are currently online. 
\item We assume that $N_s$ is sufficiently large. Small values for $N_s$,
experimentally determined to be less than about 10, lead to a large variety of
message classes that we have not yet simplified. Practical values of $N_s$ are
easily sufficient to justify this simplification.
\end{itemize}
The simplifications allow the multiple source case to be a direct result of the
single source case. 

In our model, arriving messages are randomly assigned to one of the $N_s$ peers
and so if $N_s$ is sufficiently large then a new message arriving at a peer has
a probability of $\lambda/(\mu+\lambda)$ of arriving on a online peer (these
are the $1$ messages) and it arrives on an offline peer otherwise (these are
the $0$ messages).  Our analysis however takes into account that some source
peers may not be online at some times, by allowing the possibility of $0$
messages but then ignoring them for coverage rate equations. The $0$ messages
also never enter the buffer and so they do not reduce the coverage that way
either. 

Under the aforementioned circumstances, we consider $1$ messages in different
classes determined by the number of subsequent $0$ messages that occur before
the next $1$ message, to make $i$ messages in total. Clearly the probability
for each class is a Bernoulli trial, and the time for $i$ messages to occur is
given by the Erlang distribution similarly to Eq. \ref{erlang-cov}. We arrive
at a coverage for the unit buffer size:
\[
\sum_{i=1}^{\infty}\frac{\mu^{i-1}\,\lambda}{(\lambda+\mu)^{i}}\,\widehat{C}_{1-i}.
\] 
Note that the coverage of a message in the multisource case is the same for all
of the sources. If we are considering a buffer of size $k$ then we need to
consider the arrival of $k$ messages that are inter-dispersed among the $i\geq
k$ messages and we see that there are ${i-1} \choose {k-1}$ combinations.
Therefore we obtain:
\[
\sum_{i=k}^{\infty}{{i-1} \choose {k-1}} \frac{\mu^{i-k}\,\lambda}{(\lambda+\mu)^{i}}\,\widehat{C}_{1-i}.
\] 

\begin{figure}[h]
\psfrag{alpha}{$\alpha$}
\psfrag{l=0.5}{$\lambda = 0.5$}
\psfrag{l=1.0}{$\lambda = 1.0$}
\psfrag{l=2.0}{$\lambda = 2.0$}
\psfrag{coverage}{coverage}
\includegraphics[width=\linewidth]{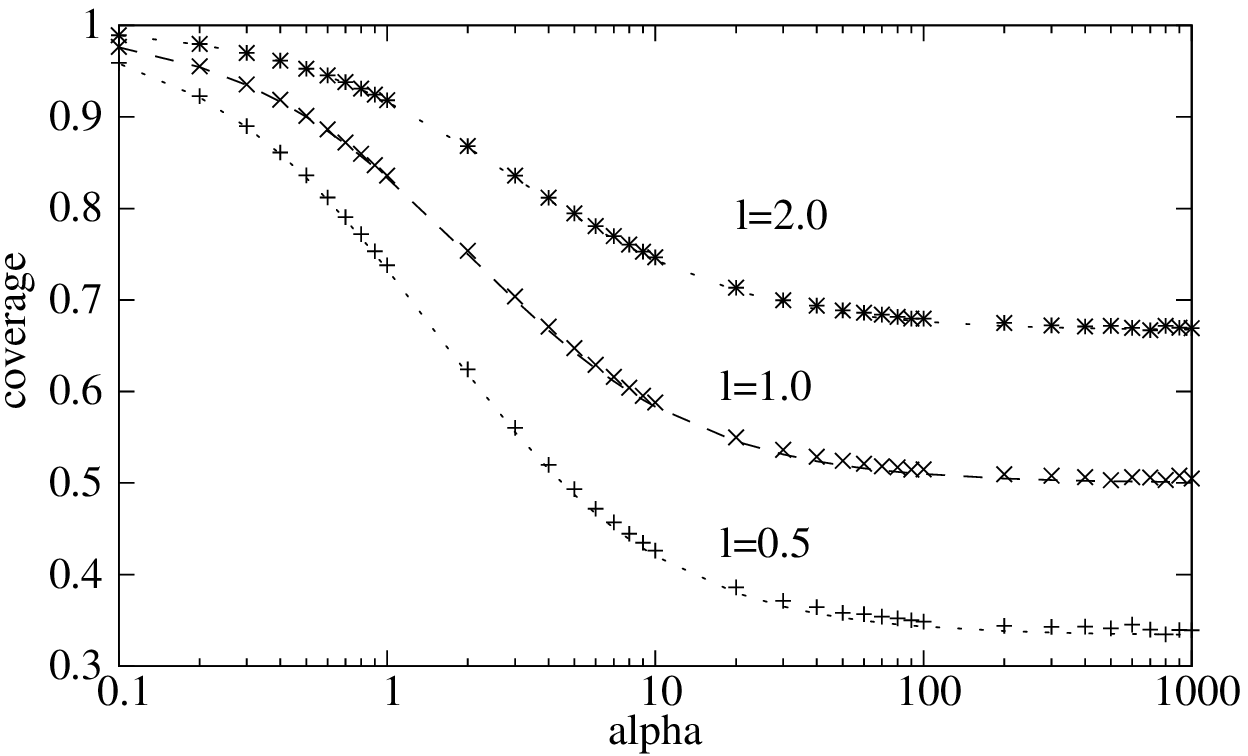}
\caption{Averaged simulation runs (points) and theoretical curves (solid lines)
for $\widehat{C}$ versus $\alpha$ with ${\mu=1.0}$, ${\lambda=1.0}$, and $k=1$.}
\label{testk1}
\end{figure}

Figure \ref{testk1} shows the coverage over a large range of $\alpha$ for
$N_s=N=100$ nodes and $k=1$. Different values of $\lambda$ are shown. The
coverages in this section cannot in general be compared with the coverages of
the previous section because the previous section included $0$ messages that
reduced the coverage. Note that the theoretical coverage can be seen,
especially for $\lambda=0.5$, to be slightly lower than the simulation. This is
due to the assumption that $N_s$ is sufficiently large. The assumption becomes
worse as $\lambda$ becomes smaller because the effective number of source peers
that are online reduces.

\begin{figure}[h]
\psfrag{l=0.5}{$\lambda = 0.5$}
\psfrag{l=1.0}{$\lambda = 1.0$}
\psfrag{l=2.0}{$\lambda = 2.0$}
\includegraphics[width=\linewidth]{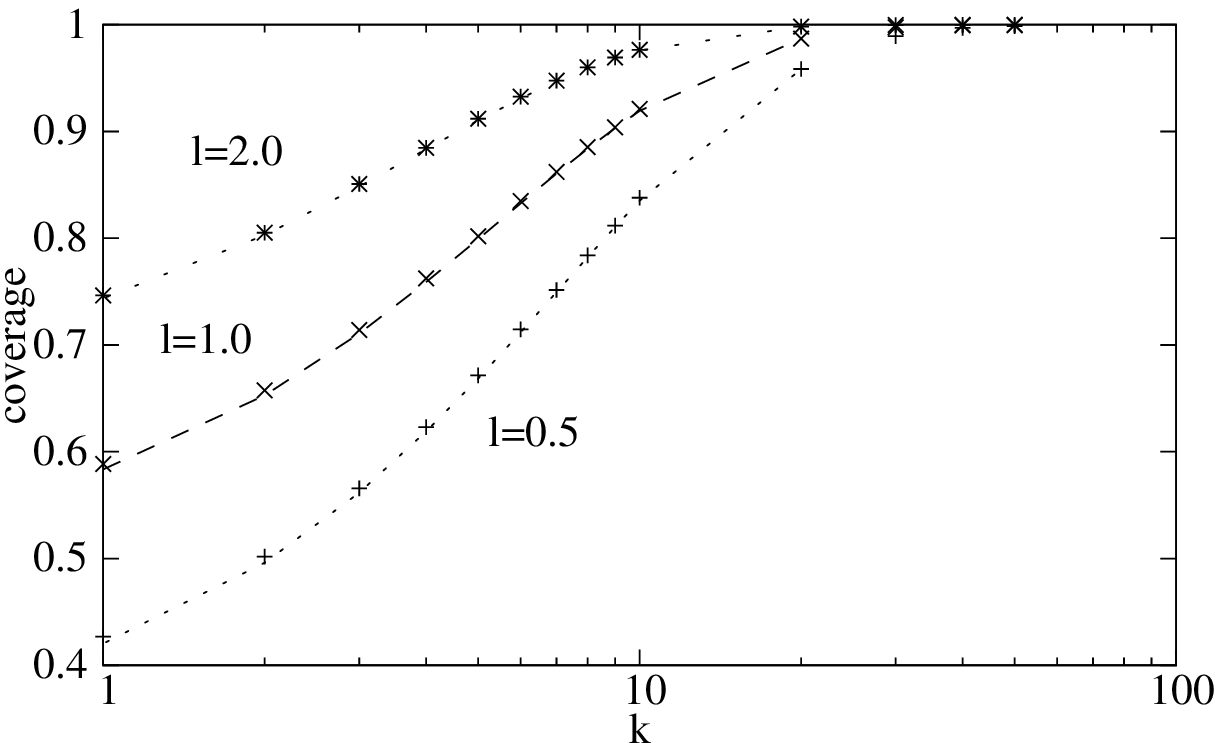}
\caption{Average simulation runs (points) and theoretical curves (solid lines) for
coverage versus $k$ for $N_s=N=100$, $\mu=1$, $\alpha=10$ and $\lambda$ as shown.}
\label{kbuftest}
\end{figure}

The increase in coverage versus $k$ is shown in Figure \ref{kbuftest}.
Numerical computation of the theoretical values became inaccurate beyond
$k=20$. Note that the chart is for the case when $\alpha=10$. Also, again the
theory slightly undershoots the simulation result as $\lambda$ becomes smaller.

Be aware that the value for $\alpha$ shown in Figures \ref{testk1} and
\ref{kbuftest} are ``net" or ``total" messages rates. Each peer is providing
messages at a rate of only $\alpha/N_s$. Therefore for $N=N_s=100$ and
$\alpha=100$, the \emph{effective} rate of a message stream from a given peer
is only $1$ per second. Thus, the coverage rate for that peer is similarly
less, even though the coverage of the messages is the same. In other words,
considering Figure \ref{testk1}, if we look at the coverage at $\alpha=100$ we
should consider the rate of messages from a single peer to be only $1$ and the
coverage rate is then 100 times worse than the single source case. Increasing
buffer size can allow us to increase $\alpha$ without sacrificing coverage and
hence to maintain a steady effective message rate per peer.

\begin{figure}[h]
\psfrag{c=0.8}{$\bar{C} = 0.8$}
\psfrag{c=0.9}{$\bar{C} = 0.9$}
\psfrag{alpha}{$\alpha$}
\includegraphics[width=\linewidth]{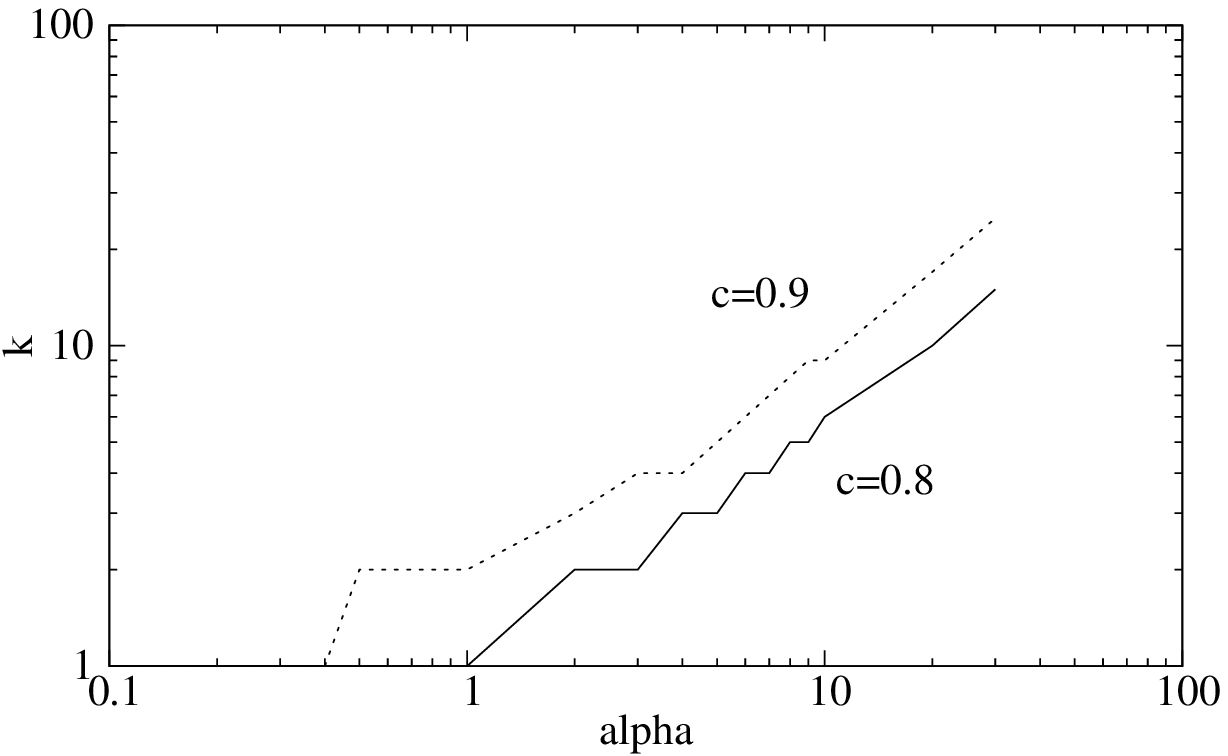}
\caption{Theoretical smallest value of $k$ that gives $\bar{C}$ as shown, for $\mu=\lambda=1$
over the $\alpha$ range.}
\label{bestk}
\end{figure}

Figure \ref{bestk} shows the theoretical smallest value of $k$ that maintains a
given coverage as $\alpha$ increases. Clearly the buffer requirements increase
proportionally to the message rate.

\section{Conclusion}
\label{conclusion}

We have proposed the Instantaneous Message Exchange (IME) model as a fundamental
approach for analyzing the affect of churn on streaming message rates. We
derived very accurate equations to describe the behavior of the P2P system and
we showed how the equations can be used in various ways to determine good
system settings. E.g., we can choose appropriate limitations on message
transmission rates with respect to churn (or vice versa) in order to achieve
high message throughput. We can also see how buffer size enhances the message
throughput and how the number of source peers affects these relationships.

In our analysis we have attempted to provide the most accurate descriptions of
the system, over all ranges of parameters. In a number of cases the equations
are complicated, including three or four complicated terms that are significant
at different ranges of the parameters. The IME model was instrumental in
allowing us to derive these equations. It remains to be seen whether we can
maintain the accuracy of the theoretical work while moving towards a
non-instantaneous model. 

Future work includes: (\emph{i}) including peer bandwidth and network delay
limitations, (\emph{ii}) examining more general communication patterns,
(\emph{iii}) using a Pareto distribution or other more suitable distribution
from trace data and (\emph{iv}) developing algorithms that reach the maximum
coverage rates.

\appendix[Simulation tool]
\label{simtool}

We developed a basic simulation tool to test our models against. In a nutshell,
the simulator is event based and maintains states for $N$ peers, including
whether each one is online or offline, which messages it has received or sent,
etc. The events of the simulator include message generation events (i.e. a peer
generates a message), and transition events from online to offline and vice
versa.  Events for the transmission of a message from one peer to another are
not part of the IME model and are therefore not required in the simulation.
Message transmission is purely a consequence of message generation and peer
online/offline states and transitions between these states.

The simulation begins by assigning each peer as online or offline, with
probability $\lambda/(\lambda+\mu)$ and $\mu/(\lambda+\mu)$ resp. In the single
message case the simulation then chooses a peer at random and runs until the
message has been received by all peers. For the message stream case the
simulation continues to run until a given number of messages have been
generated. A sufficiently large number of messages need to be generated in
order for the observed results to be independent of the starting configuration
of online/offline peers, in other words depending on $\mu$, $\lambda$ and
$\alpha$. An appropriate number of messages was set experimentally. For the
multiple source case the simulation randomly chooses $N_s$ peers as source
peers. 

Event times are real numbers and the simulator orders all pending events
and processes one event at a time.
Following is the brief explanation of what happens after each event:
\begin{enumerate}
\item Message generation:
\begin{enumerate}
\item Peer was online:
\begin{itemize}
\item Send new message to all of the online peers.
\item Compute next message generating event for that peer.
\end{itemize}
\item Peer was offline:
\begin{itemize}
\item Compute next message generating event for that peer.
\end{itemize}
\end{enumerate}
\item Changing from online to offline:
\begin{itemize}
\item Schedule the next time to change to online.
\end{itemize}
\item Changing from offline to online:
\begin{itemize}
\item Merge buffers with all online peers to produce the buffer with newest
messages; see notes below.
\item Send that buffer to all of the online peers.
\item Schedule the next time to change to offline. 
\end{itemize}
\end{enumerate}

Notes:
\begin{itemize}
\item An invariant of the simulation is that at any point in time the
buffers of the online peers are identical. This is a consequence of the IME
model. If bandwidth or latency for message transmission were taken into account
then the invariant would be broken.
\item Each message is assigned a number such that message $i$ is newer than
message $j$ only if $j<i$, i.e. message $i$ was generated later than message
$j$.  A buffer is always sorted by the message number.  The buffer is changed
due to: (\emph{i}) new message coming into the network; or (\emph{ii}) a node
that has newer messages than some of the messages in the online peers' buffers,
enters the network. In the first case the newer message is appended to the
buffer when the current buffer size is less than $k$; if the current buffer
size is equal to $k$ then the oldest message of the buffer is pushed out.  In
the second case the buffers of the just arrived peer and any online peer are
merged so that the merged buffer has the $k$ newest messages of the union of
those two older buffers. Each of the online peers then has the merged buffer.
\end{itemize}

\bibliographystyle{abbrv}
\bibliography{ton}

\begin{thebibliography}{10}

\bibitem{bail75}
N.~T.~J. Bailey.
\newblock {\em The Mathematical Theory of Infectious Diseases and its
  Applications}.
\newblock Griffin, 2nd edition, 1975.

\bibitem{karl}
A.~Datta, S.~Quarteroni, and K.~Aberer.
\newblock Autonomous gossiping: A self-organizing epidemic algorithm for
  selective information dissemination in wireless mobile ad-hoc networks.
\newblock Technical Report EPFL Technical Report IC-2004-07, Swiss Federal
  Institute of Technology, 2004.

\bibitem{citeulike:392472}
P.~T. Eugster and R.~Guerraoui.
\newblock Probabilistic multicast.
\newblock In {\em Proceedings of the International Conference on Dependable
  Systems and Networks}, pages 313--322, 2002.

\bibitem{citeulike:292941}
P.~T. Eugster, R.~Guerraoui, A.~M. Kermarrec, and L.~Massoulie.
\newblock Epidemic information dissemination in distributed systems.
\newblock {\em Computer}, 37(5):60--67, 2004.

\bibitem{karp}
R.~Karp, C.~Schindelhauer, S.~Shenker, and B.~{V\"ocking}.
\newblock Randomized rumor spreading.
\newblock In {\em Proceedings of the 41st Annual Symposium on Foundations of
  Computer Science - FOCS}, pages 565--574. IEEE Computer Society, 2000.

\bibitem{DBLP:conf/sigmetrics/Lindemann05}
C.~Lindemann and O.~P. Waldhorst.
\newblock Modeling epidemic information dissemination on mobile devices with
  finite buffers.
\newblock In {\em SIGMETRICS}, pages 121--132, 2005.

\bibitem{constand}
C.~X. Mavromoustakis and H.~D. Karatza.
\newblock Segmented file sharing with recursive epidemic placement policy for
  reliability in mobile peer-to-peer devices.
\newblock In {\em Proceedings of the 13th IEEE International Symposium on
  Modeling, Analysis, and Simulation of Computer and Telecommunication Systems
  (MASCOTS)}, page 8 pages, 2005.

\bibitem{papa}
M.~Papadopouli and H.~Schulzrinne.
\newblock Effects of power conservation, wireless coverage and cooperation on
  data dissemination among mobile devices.
\newblock In {\em MobiHoc '01: Proceedings of the 2nd ACM international
  symposium on Mobile ad hoc networking and computing}, pages 117--127, 2001.

\bibitem{werner}
W.~Vogels, R.~van Renesse, and K.~Birman.
\newblock The power of epidemics: Robust communication for large-scale
  distributed systems.
\newblock {\em ACM SIGCOMM Computer Communications Review}, 33(1):131--135,
  2003.

\bibitem{yao}
Z.~Yao, D.~Leonard, X.~Wang, and D.~Loguinov.
\newblock Modeling heterogeneous user churn and local resilience of
  unstructured p2p networks.
\newblock In {\em Proceedings of the 2006 14th IEEE International Conference on
  Network Protocols}, pages 32--41, 2006.

\end{thebibliography}

\end{document}